\documentclass[10pt, paper=a4, UKenglish, dvipsnames]{article}
\usepackage{graphicx}
\usepackage{tikz}

\def\Title#1{\begin{center} {\Large #1 } \end{center}}
\def\Author#1{\begin{center}{ \sc #1} \end{center}}
\def\Address#1{\begin{center}{ \it #1} \end{center}}

\newcommand\pubblock{\rightline{\begin{tabular}{l} \pubnumber\\
         \pubdate  \end{tabular}}}

\newenvironment{Abstract}{\begin{quotation} \begin{center} 
             \large ABSTRACT \end{center}\bigskip 
      \begin{center}\begin{large}}{\end{large}\end{center} \end{quotation}}

\newenvironment{Presented}{\begin{quotation} \begin{center} 
             PRESENTED AT\end{center}\bigskip 
      \begin{center}\begin{large}}{\end{large}\end{center} \end{quotation}}
      
\def\papercopyright{\the\year\ CERN for the benefit of the LHCb collaboration} 
\def\paperlicence{CC BY 4.0 licence}
\def\paperlicenceurl{https://creativecommons.org/licenses/by/4.0/}

\def\Acknowledgements{\bigskip  \bigskip \begin{center} \begin{large}
      \bf ACKNOWLEDGEMENTS \end{large}\end{center}}

\usepackage{microtype}
\usepackage{lineno}  
\usepackage{xspace} 

\usepackage{graphicx}  
\graphicspath{{./figs/}} 

\usepackage{amsmath} 
\usepackage{amssymb}
\usepackage{amsfonts}
\usepackage{upgreek} 
\usepackage{anyfontsize}

\newcommand*\patchAmsMathEnvironmentForLineno[1]{%
\expandafter\let\csname old#1\expandafter\endcsname\csname #1\endcsname
\expandafter\let\csname oldend#1\expandafter\endcsname\csname
end#1\endcsname
 \renewenvironment{#1}%
   {\linenomath\csname old#1\endcsname}%
   {\csname oldend#1\endcsname\endlinenomath}%
}
\newcommand*\patchBothAmsMathEnvironmentsForLineno[1]{%
  \patchAmsMathEnvironmentForLineno{#1}%
  \patchAmsMathEnvironmentForLineno{#1*}%
}
\AtBeginDocument{%
\patchBothAmsMathEnvironmentsForLineno{equation}%
\patchBothAmsMathEnvironmentsForLineno{align}%
\patchBothAmsMathEnvironmentsForLineno{flalign}%
\patchBothAmsMathEnvironmentsForLineno{alignat}%
\patchBothAmsMathEnvironmentsForLineno{gather}%
\patchBothAmsMathEnvironmentsForLineno{multline}%
\patchBothAmsMathEnvironmentsForLineno{eqnarray}%
}

\usepackage{hyperxmp}

\usepackage[colorinlistoftodos,textsize=scriptsize]{todonotes}

\usepackage{ifthen} 
\newboolean{uprightparticles}
\setboolean{uprightparticles}{false} 
\usepackage{xspace} 
\usepackage{upgreek}

\def\lhcb   {\mbox{LHCb}\xspace}

\def\lhc    {\mbox{LHC}\xspace}

\def\velo   {VELO\xspace}
\def\rich   {RICH\xspace}

\def\MagUp {\mbox{\em Mag\kern -0.05em Up}\xspace}

\def\hlttwo {HLT2\xspace}

\ifthenelse{\boolean{uprightparticles}}%
{

 \def\PDelta      {\ensuremath{\Delta}\xspace}                 
 \def\PXi         {\ensuremath{\Xi}\xspace}                 
 \def\PLambda     {\ensuremath{\Lambda}\xspace}                 
 \def\PSigma      {\ensuremath{\Sigma}\xspace}                 
 \def\POmega      {\ensuremath{\Omega}\xspace}                 
 \def\PUpsilon    {\ensuremath{\Upsilon}\xspace}
 \let\oldPi\Pi
 \def\PPi         {\ensuremath{\oldPi}\xspace}

 \def\PB      {\ensuremath{\mathrm{B}}\xspace}                 
                  
 \def\PD      {\ensuremath{\mathrm{D}}\xspace}

 \def\PK      {\ensuremath{\mathrm{K}}\xspace}

 \def\Pi      {\ensuremath{\mathrm{i}}\xspace}

 \def\Pp      {\ensuremath{\mathrm{p}}\xspace}

 \def\Ps      {\ensuremath{\mathrm{s}}\xspace}

 \def\thebaroffset{0.0em}
}
{

 \mathchardef\PDelta="7101
 \mathchardef\PXi="7104
 \mathchardef\PLambda="7103
 \mathchardef\PSigma="7106
 \mathchardef\POmega="710A
 \mathchardef\PUpsilon="7107
 \mathchardef\PPi="7105
                  
 \def\PB      {\ensuremath{B}\xspace}                 
                  
 \def\PD      {\ensuremath{D}\xspace}

 \def\PK      {\ensuremath{K}\xspace}

 \def\Pi      {\ensuremath{i}\xspace}

 \def\Pp      {\ensuremath{p}\xspace}

 \def\Ps      {\ensuremath{s}\xspace}

 \def\thebaroffset{0.18em}
}
\newcommand{\offsetoverline}[2][\thebaroffset]{\kern #1\overline{\kern -#1 #2}}%

\makeatletter
\ifcase \@ptsize \relax
  \newcommand{\miniscule}{\@setfontsize\miniscule{4}{5}}
\or
  \newcommand{\miniscule}{\@setfontsize\miniscule{5}{6}}
\or
  \newcommand{\miniscule}{\@setfontsize\miniscule{5}{6}}
\fi
\makeatother

\DeclareRobustCommand{\optbar}[1]{\shortstack{{\miniscule (\rule[.5ex]{1.25em}{.18mm})}
  \\ [-.7ex] $#1$}}

\def\squark    {{\ensuremath{\Ps}}\xspace}

\def\KorKbar {\kern \thebaroffset\optbar{\kern -\thebaroffset \PK}{}\xspace}

\def\D       {{\ensuremath{\PD}}\xspace}

\def\DorDbar {\kern \thebaroffset\optbar{\kern -\thebaroffset \PD}\xspace}

\def\Dp      {{\ensuremath{\D^+}}\xspace}
\def\Dm      {{\ensuremath{\D^-}}\xspace}

\def\DpDm    {\ensuremath{\Dp {\kern -0.16em \Dm}}\xspace}

\def\B       {{\ensuremath{\PB}}\xspace}

\def\BorBbar {\kern \thebaroffset\optbar{\kern -\thebaroffset \PB}\xspace}

\def\Bd      {{\ensuremath{\B^0}}\xspace}

\def\BdorBdbar {\kern \thebaroffset\optbar{\kern -\thebaroffset \Bd}\xspace}

\def\Bs      {{\ensuremath{\B^0_\squark}}\xspace}

\def\BsorBsbar {\kern \thebaroffset\optbar{\kern -\thebaroffset \Bs}\xspace}

\def\Y#1S{\ensuremath{\PUpsilon{(#1S)}}\xspace}

\def\proton      {{\ensuremath{\Pp}}\xspace}

\def\LorLbar     {\kern \thebaroffset\optbar{\kern -\thebaroffset \PLambda}\xspace}

\def\order   {{\ensuremath{\mathcal{O}}}\xspace}

\def\AT#1     {\ensuremath{A_{\mathrm{T}}^{#1}}\xspace}           

\def\C#1      {\ensuremath{\mathcal{C}_{#1}}\xspace}                       
\def\Cp#1     {\ensuremath{\mathcal{C}_{#1}^{'}}\xspace}                    
\def\Ceff#1   {\ensuremath{\mathcal{C}_{#1}^{\mathrm{(eff)}}}\xspace}        
\def\Cpeff#1  {\ensuremath{\mathcal{C}_{#1}^{'\mathrm{(eff)}}}\xspace}       
\def\Ope#1    {\ensuremath{\mathcal{O}_{#1}}\xspace}                       
\def\Opep#1   {\ensuremath{\mathcal{O}_{#1}^{'}}\xspace}                    

\newcommand{\aunit}[1]{\ensuremath{\text{\,#1}}}       

\newcommand{\tev}{\aunit{Te\kern -0.1em V}\xspace}
\newcommand{\gev}{\aunit{Ge\kern -0.1em V}\xspace}
\newcommand{\mev}{\aunit{Me\kern -0.1em V}\xspace}
\newcommand{\kev}{\aunit{ke\kern -0.1em V}\xspace}
\newcommand{\ev}{\aunit{e\kern -0.1em V}\xspace}
 
\newcommand{\mevc}{\ensuremath{\aunit{Me\kern -0.1em V\!/}c}\xspace}
\newcommand{\gevc}{\ensuremath{\aunit{Ge\kern -0.1em V\!/}c}\xspace}
\newcommand{\mevcc}{\ensuremath{\aunit{Me\kern -0.1em V\!/}c^2}\xspace}
\newcommand{\gevcc}{\ensuremath{\aunit{Ge\kern -0.1em V\!/}c^2}\xspace}

\def\cm   {\aunit{cm}\xspace}

\def\sec  {\ensuremath{\aunit{s}}\xspace}

\def\mhz  {\ensuremath{\aunit{MHz}}\xspace}

\def\order{{\ensuremath{\mathcal{O}}}\xspace}

\def\gsim{{~\raise.15em\hbox{$>$}\kern-.85em
          \lower.35em\hbox{$\sim$}~}\xspace}
\def\lsim{{~\raise.15em\hbox{$<$}\kern-.85em
          \lower.35em\hbox{$\sim$}~}\xspace}

\newcommand{\lum} {\ensuremath{\mathcal{L}}\xspace}

\def\tell1  {TELL1\xspace}
\def\ukl1   {UKL1\xspace}

\newcommand{\eg}{\mbox{\itshape e.g.}\xspace}
\newcommand{\ie}{\mbox{\itshape i.e.}\xspace}

\usepackage{cite} 

\usepackage{color}
\usepackage{lineno}
\usepackage{subfig}
\usepackage{hyperref}

\newcommand\pubnumber{LHCb-PROC-2022-009}

\newcommand\pubdate{October 26, 2022}

\def\affiliation{
On behalf of the \lhcb collaboration, \\
Physikalisches Institut \\
Heidelberg University, Germany}

\newcommand{\conference}{Connecting the Dots Workshop (CTD 2022)\\
May 31 - June 2, 2022}

\usepackage{fancyhdr}
\definecolor{mygrey}{RGB}{105,105,105}
\begin{document}


\large
\begin{titlepage}
\pubblock

\vfill
\Title{LHCb's Forward Tracking algorithm for the Run 3 CPU-based online track-reconstruction sequence}
\vfill

\Author{Paul André Günther}
\Address{\affiliation}
\vfill

\begin{Abstract}
In Run 3 of the \lhc, the \lhcb experiment faces very high data rates containing beauty and charm hadron decays. Thus the task of the trigger is not to select any beauty and charm events, but to select those containing decays interesting for the \lhcb physics programme. \lhcb has therefore implemented a real-time data processing strategy to trigger directly on reconstructed events. The first stage of the purely software-based trigger is implemented on GPUs performing a partial event reconstruction. In the second stage of the software trigger, the full, offline-quality event reconstruction is performed on CPUs, with a crucial part being track reconstruction, balancing track finding efficiency, fake track rate and event throughput. \lhcb's CPU-based track reconstruction sequence for Run 3 is presented, highlighting the "Forward Tracking", which is the algorithm that reconstructs trajectories of charged particles traversing all of \lhcb's tracking detectors. To meet event throughput requirements, the "Forward Tracking" uses SIMD instructions in several core parts of the algorithm, such as the Hough transform and the cluster search. These changes led to an improvement of the algorithm's event throughput by 60\%.
\end{Abstract}

\vfill

\begin{Presented}
\conference
\end{Presented}
\vfill
{\footnotesize
\centerline{\copyright~\papercopyright. \href{\paperlicenceurl}{\paperlicence}.}}
\end{titlepage}
\def\thefootnote{\fnsymbol{footnote}}
\setcounter{footnote}{0}
%

\normalsize 


\section{Introduction}
\label{intro}

During Run 3 of the \lhc, \lhcb will take data with new tracking detectors, new readout electronics and a new, purely software-based trigger \cite{LHCb:2012doh,CERN-LHCC-2014-016}. These upgrades to the \lhcb experiment are necessary to efficiently process the data from 30\mhz of non-empty \proton\proton bunch crossings at the \lhcb design luminosity of $\lum=2\times10^{33}\cm^{-2}\sec^{-1}$ producing millions of beauty and charm hadrons per second. At these signal rates a hardware trigger is not anymore able to efficiently reduce the amount of data recorded as its bandwidth is already fully saturated by hadronic \B decays \cite{CERN-LHCC-2014-016,Piucci:2017kih}. This problem is solved by a purely software-based trigger system that analyses and selects events in real time, \ie performing the full, offline-quality event reconstruction. The most crucial part of the reconstruction sequence is the charged track reconstruction. The Run 3 \lhcb detector employs three tracking detector technologies, a hybrid pixel detector (\velo, Vertex Locator) surrounding the \proton\proton interaction region \cite{Collaboration:2013kdk}, a silicon microstrip detector (UT, Upstream Tracker) \cite{Collaboration:2014zyt} placed upstream of a dipole magnet and three stations of large-area scintillating fibre detectors (SciFi) \cite{Collaboration:2014zyt} located downstream of the magnet. The whole \lhcb detector is a single-arm forward spectrometer covering the pseudorapidity range $2<\eta<5$. The software trigger is split into two parts, the first stage is the \textit{High Level Trigger 1} which performs a partial event reconstruction on GPUs \cite{Aaij:2019zbu,LHCb:2021kxm,PROC-CTD2022-28,PROC-CTD2022-33}, the second stage, called \textit{High Level Trigger 2} (\hlttwo), reconstructs the full event using CPUs and selects decay candidates which are then sent to permanent storage for further analysis \cite{CERN-LHCC-2018-014}. The full \hlttwo event reconstruction includes charged track reconstruction, track parameter estimation using a Kalman filter, hadron, electron and muon identification and photon reconstruction. The \hlttwo charged track reconstruction sequence is outlined in Sec.~\ref{sec:reco}. One of the algorithms in this sequence is the \textit{Forward Tracking}, highlighted in Sec.~\ref{sec:forward}.

\section{CPU-based charged track reconstruction in \hlttwo}
\label{sec:reco}

\lhcb defines five different track types described in Ref.~\cite{Li:2021oga}. Among these are \velo, T and Long tracks. \velo and T tracks have hits only in the \velo and SciFi detectors, respectively, while Long tracks traverse the whole tracking system and are required to have hits in both, the \velo and SciFi detectors, and can optionally have hits in the UT detector. Because they use all tracking detectors, Long tracks have the best momentum resolution and are therefore most used for physics analysis. 

\begin{figure}[!htb]
    \centering
    \subfloat[]{\usetikzlibrary{shapes, arrows.meta, positioning}
\begin{tikzpicture}[font=\small]
	\node[draw,
	minimum width=2cm,
	color=DarkOrchid,
	align=left,
	minimum height=1cm] (velotracking) {VELO Tracking
	};
	
	\node[draw,
	trapezium,
	below=of velotracking,
	trapezium left angle = 120,
	trapezium right angle = 120,
	color=DarkOrchid,
	trapezium stretches] (velotracks) {VELO Tracks};
	
	\node[draw,
	right=of velotracking,
	minimum width=2cm,
	color=OliveGreen,
	align=left,
	minimum height=1cm] (hybridseeding) {Hybrid Seeding
	};
	
	\node[draw,
	trapezium,
	below=of hybridseeding,
	color=OliveGreen,
	trapezium left angle = 120,
	trapezium right angle = 120,
	trapezium stretches] (seedtracks) {T Tracks};
	
	
	\node[draw,
	below=of seedtracks,
	minimum width=2cm,
	color=Mahogany,
	minimum height=1cm] (matching) {Matching};
	
			\node[draw,
	trapezium,
	below=of velotracks,
	trapezium left angle = 120,
	trapezium right angle = 120,
	color=DarkOrchid,
	trapezium stretches] (unmatched) {residual VELO Tracks};
	
	\node[draw,
	below=of unmatched,
	minimum width=2cm,
	color=OrangeRed,
	minimum height=1cm] (forwardtracking) {Forward Tracking};
	
	\node[draw,
	trapezium,
	below =of matching,
	trapezium left angle = 120,
	trapezium right angle = 120,
	color=Maroon,
	trapezium stretches] (longtracks) {Long Tracks};
	
	\draw[-latex, line width=0.3mm] (velotracking) edge (velotracks)
	(velotracks) edge (matching)
	(matching) edge (unmatched)
	(unmatched) edge (forwardtracking)
	(forwardtracking) edge (longtracks)
	(hybridseeding) edge (seedtracks)
	(seedtracks) edge (matching)
	(matching) edge (longtracks);
	
\end{tikzpicture}}
    \qquad
    \subfloat[]{\includegraphics[width=0.5\linewidth]{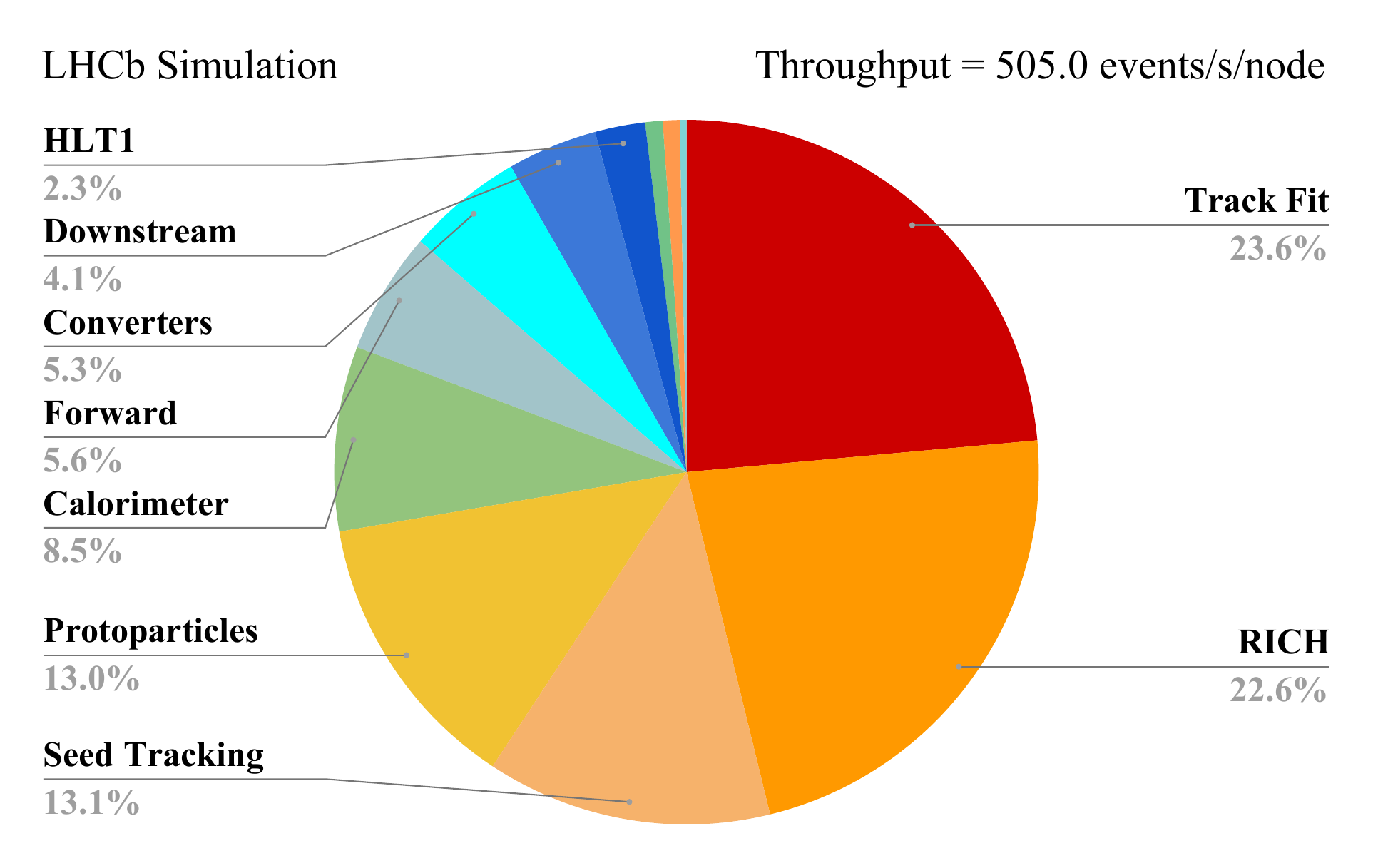}}
    \caption{The Run 3 Long Track reconstruction sequence is shown in (a). Colours indicate that objects belong together. The throughput contribution of different components in the \hlttwo reconstruction sequence is shown in (b) \cite{LHCb-Figure-2022-005}. \rich, Calorimeter and Protoparticles refer to particle identification, the Track Fit is a Kalman filter. The Hybrid Seeding algorithm is called Seed Tracking here.}
    \label{fig:reco_sequence}
\end{figure}
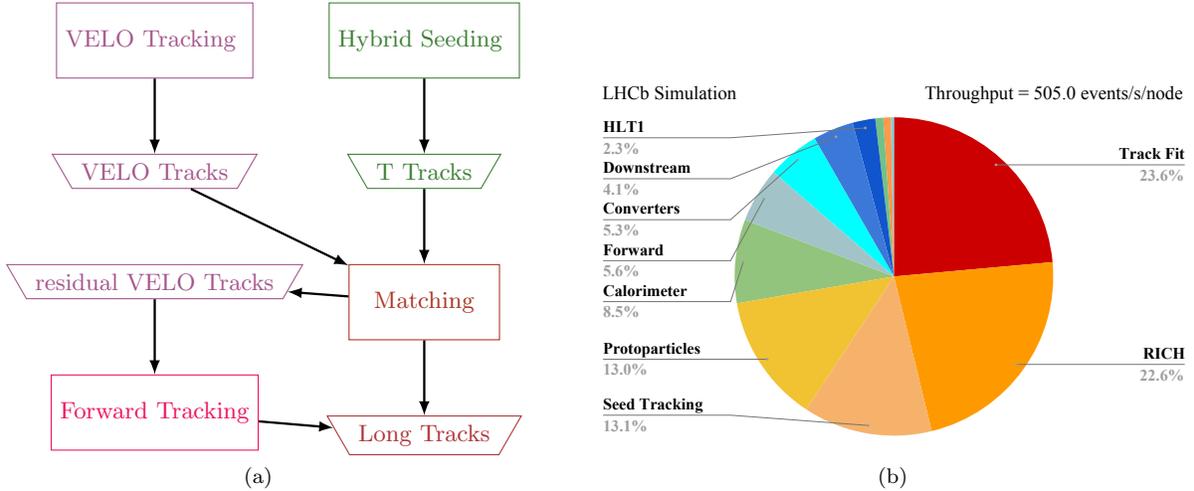

In \hlttwo, two independent algorithms reconstruct Long tracks, the \textit{Forward Tracking} and the \textit{Matching}, each of which employs a different track reconstruction method. The Forward Tracking and Matching both take \velo tracks as input, the reconstruction of which is done by the \velo Tracking described in Ref.~\cite{Hennequin:2019itm}. The Matching further takes T tracks, reconstructed by the Hybrid Seeding algorithm described in Ref.~\cite{Quagliani:2017umh}, and matches them to \velo tracks which consequently become Long tracks. To find the correct combinations of \velo and T tracks, the Matching evaluates a multilayer perceptron trained on simulation and decides whether there is a match depending on the response of the neural network. The Forward Tracking is described in Sec.~\ref{sec:forward}. Both algorithms are designed to find as many Long tracks as possible, which leads to a large overlap between the two sets of found Long tracks. In Run 2, this redundancy was accepted and resolved by selecting Long tracks according to their quality after applying a Kalman filter. This procedure increased the reconstruction efficiency \cite{LHCb:2018zdd}. In Run 3, however, the timing constraints posed by the real-time analysis trigger strategy favour avoiding redundancy. This is achieved by the reconstruction sequence shown in Fig.~\ref{fig:reco_sequence}a. First the \velo tracking and Hybrid Seeding are run. The Matching then combines \velo tracks with T tracks. Residual \velo tracks, that could not be matched, are given to the Forward Tracking as input, which tries to create a Long Track using only SciFi hits that are not already taken into account by a Long Track found by the Matching. 

The event throughput of the \hlttwo reconstruction sequence is shown in Fig.~\ref{fig:reco_sequence}b, including contributions from the Kalman filter and particle identification. Avoiding the aforementioned redundancy in the Long Track reconstruction, the \hlttwo reconstruction sequence reaches more than $500$ events per second and computing node, with the Forward Tracking not being a throughput-dominating component. More performance metrics such as reconstruction efficiencies of the \hlttwo charged track reconstruction can be found in Ref.~\cite{LHCb-Figure-2021-003}.

\section{\hlttwo Forward Tracking}
\label{sec:forward}

The goal of the Forward Tracking is to find a forward extension of a given \velo track in the SciFi tracker and to estimate a preliminary\footnote{The best momentum estimate is obtained by applying a Kalman filter to the track found bound by the Forward Tracking, using the preliminary momentum estimate as input.} momentum of this Long Track. The SciFi detector consists of three stations, each with four layers of scintillating fibre modules. Therefore, the forward extension is a set of ten to twelve SciFi hits, with each hit belonging to a different layer of the SciFi detector. The hits the Forward Tracking algorithm searches for form a slightly curved trajectory that is compatible with originating from a \velo track. The line is curved in the $xz$- and $yz$-projection\footnote{The \lhcb coordinate system is a right-handed system with positive $z$ running along the beamline from the interaction point into the detector and positive $y$ pointing upward.} because of fringe magnetic fields within the SciFi stations. The central problem to solve is finding the single correct combination of a \velo track with SciFi hits among the many possible extensions, or finding that no such combination is present, while keeping the reconstruction efficiency and event throughput high and the fake track fraction low. The complexity scales with the number of input tracks and the total number of hits recorded by the SciFi detector. Typical events with inelastic \proton\proton collisions lead to $\order(10^2)$ \velo tracks and 
$\order(10^3)$ SciFi hits. The Forward Tracking uses a method similar to a Hough transform \cite{Hough:1959qva} to efficiently and robustly spot hits forming the line pattern matching a \velo track. The following Sections \ref{sec:hit_sel}-\ref{sec:hough} describe important components of the algorithm performed on individual input tracks.

\subsection{SciFi hit selection}
\label{sec:hit_sel}
Starting from a state vector $(x,y,\frac{\partial x}{\partial z}, \frac{\partial y}{\partial z}, \frac{q}{p})$ containing the position, slopes and so far unknown charge $q$ and momentum $p$ at the end of the \velo, \ie a \velo track, the Forward Tracking defines a polynomial $P(\frac{\partial x}{\partial z}, \frac{\partial y}{\partial z}, p)$ parameterising the propagation of the track in the $xz$-plane through the magnetic field down to the SciFi layers. As the momentum and charge are not known yet, this parameterisation is used to calculate $x_{\mathrm{min}}$ and $x_{\mathrm{max}}$ positions for each layer assuming a minimum reconstructible track momentum $p_{\mathrm{min}}$, \eg $p_{\mathrm{min}}=1.5\gev$, and both possible charges. Only SciFi hits with $x\in[x_{min},x_{max}]$ are selected for further processing, reducing the complexity of the problem. The determination of this hit search window by a polynomial is computationally cheap and therefore preferred over numerically solving the equations of motion in the magnetic field in a real-time application.

\subsection{Simplified particle trajectory}
\label{sec:traj}
Similarly to the parameterisation used to define the hit search window (Sec.~\ref{sec:hit_sel}), the Forward Tracking defines a simplified particle trajectory by treating the magnet as an optical lens as described in Ref.~\cite{Benayoun:2002hqa}. Just like the model of light rays refracted by a thin lens, the particle's movement in the $xz$-plane through the magnetic field is modelled by a straight line that gets a kick at the centre of the magnetic field and propagates further as a straight line with a different slope. Once a single hit $(x,z)$ downstream of the magnet is taken into account, predicting the $x$ coordinate at a given $z$ position is a simple linear extrapolation within the model. Deviations from this model occur because of fringe magnetic fields that reach into the SciFi detector and are corrected for by parameterising the effect using polynomials as described in Ref.~\cite{Amhis:2014abw}.

\subsection{Hough-like transform}
\label{sec:hough}
The main component of the Forward Tracking applies a map-reduce pattern inspired by the Hough transform to sieve out sets of SciFi hits that do not form a matching extension to the \velo track. 

\begin{figure}[!htb]
    \centering
    \resizebox{\textwidth}{!}{
        \fontsize{40}{45}\selectfont
        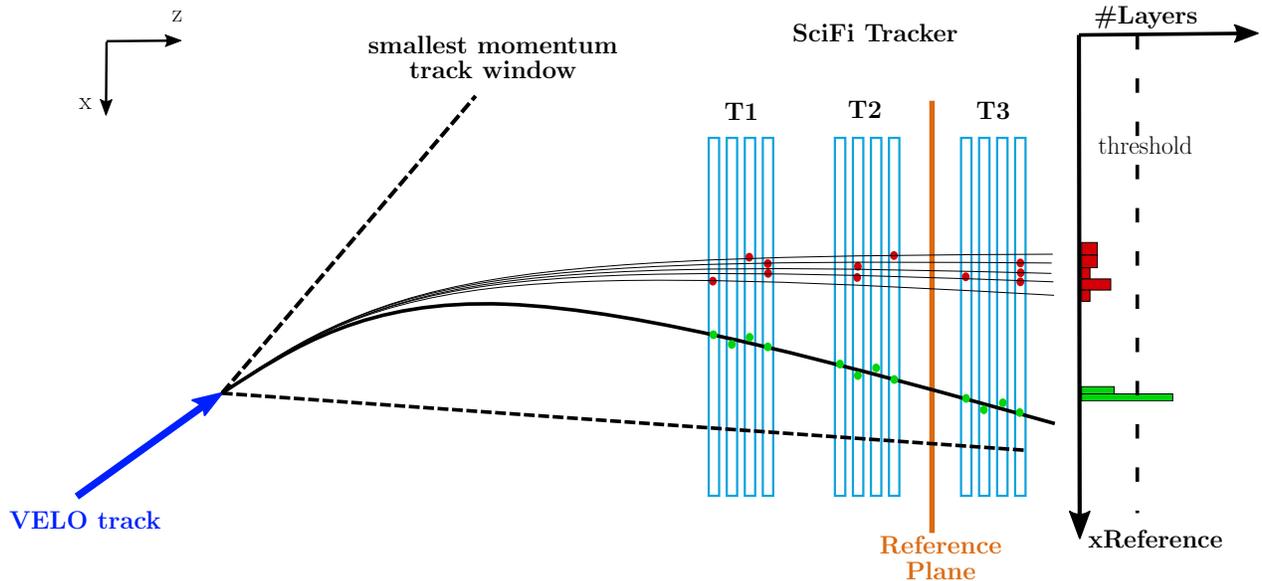
    }
    \caption{Sketch of the key components of the Forward Tracking. Starting from a \velo track (blue), a smallest-momentum hit search window is calculated (black dashed line). The $x$ positions of hits in the twelve SciFi detector layers (three stations T1, T2 and T3 with four layers each in light blue) are projected to the reference plane (orange) using a simplified track model (not shown here). The rightmost part of the figure shows the histogram counting the number of unique SciFi detector layers depending on the projected $x$ positions. Hits belonging to the \velo track are shown in green, and other hits are in red.}
    \label{fig:hough_trf}
\end{figure}

The $x$ positions of SciFi hits selected by the search window described in Sec.~\ref{sec:hit_sel} are mapped to a reference plane at a fixed $z$ position. All hits' $x$ positions at the reference plane are calculated using the simplified trajectory introduced in Sec.~\ref{sec:traj} and filled into a histogram. The histogram counts the number of unique SciFi detector layers that are present among the hits in one bin. This way, hits that do not qualify as an extension to the \velo track form a flat distribution, while hits that match the \velo track accumulate in a few bins as depicted in Fig.~\ref{fig:hough_trf}. Subsequently, the histogram is scanned for small groups of neighbouring bins exceeding a layer-count threshold, thus reducing the large set of hits from within the search window to none, one or several small sets of hits, which become candidates for the \velo track extension. The found hit sets are then cleaned from outliers, fitted using a third-order polynomial function and further selected according to the fit result. The remaining candidates are promoted to Long tracks and their charge and momentum are estimated. 

\subsection{Event throughput optimisation}
\label{sec:opti}
Optimisation of tracking algorithms usually deals with a trade-off between event throughput, track reconstruction efficiency and fake track fraction. However, the capabilities of modern CPUs offer new opportunities to improve the throughput of an algorithm without compromising the other two metrics. 

The throughput performance of the Forward Tracking is optimised by exploiting data-level parallelism using the vector registers of the CPU. These registers contain the operands for a single CPU instruction that acts on multiple data points at once (SIMD, single instruction multiple data). For this to work efficiently, an appropriate choice of data structures is crucial. Therefore, information about SciFi detector hits is stored in a structure-of-arrays layout, \ie all $x$ positions are arranged successively in memory. This enables fast data loading to SIMD registers as well as fast storing of the computed result. The Forward Tracking mostly applies this form of parallelism in the computationally expensive map-reduce part explained in Sec.~\ref{sec:hough}. The mapping of hits' $x$ positions to the reference plane is done for multiple hits at once using single instructions and likewise the threshold scan reducing the data is performed on several histogram bins in parallel. Using the Advanced Vector Extensions 2 (AVX2), SIMD registers hold $256$ bits equal to eight single precision floating point or integer numbers that then can be processed in parallel. Adapting the Forward Tracking algorithm to efficiently use AVX2 improves its event throughput by 60\% without losses regarding reconstruction efficiency or fake track fraction. 

\subsection{Reconstruction performance}
\label{sec:perf}
\begin{figure}[!htb]
    \centering
    \subfloat[]{\includegraphics[width=0.475\linewidth]{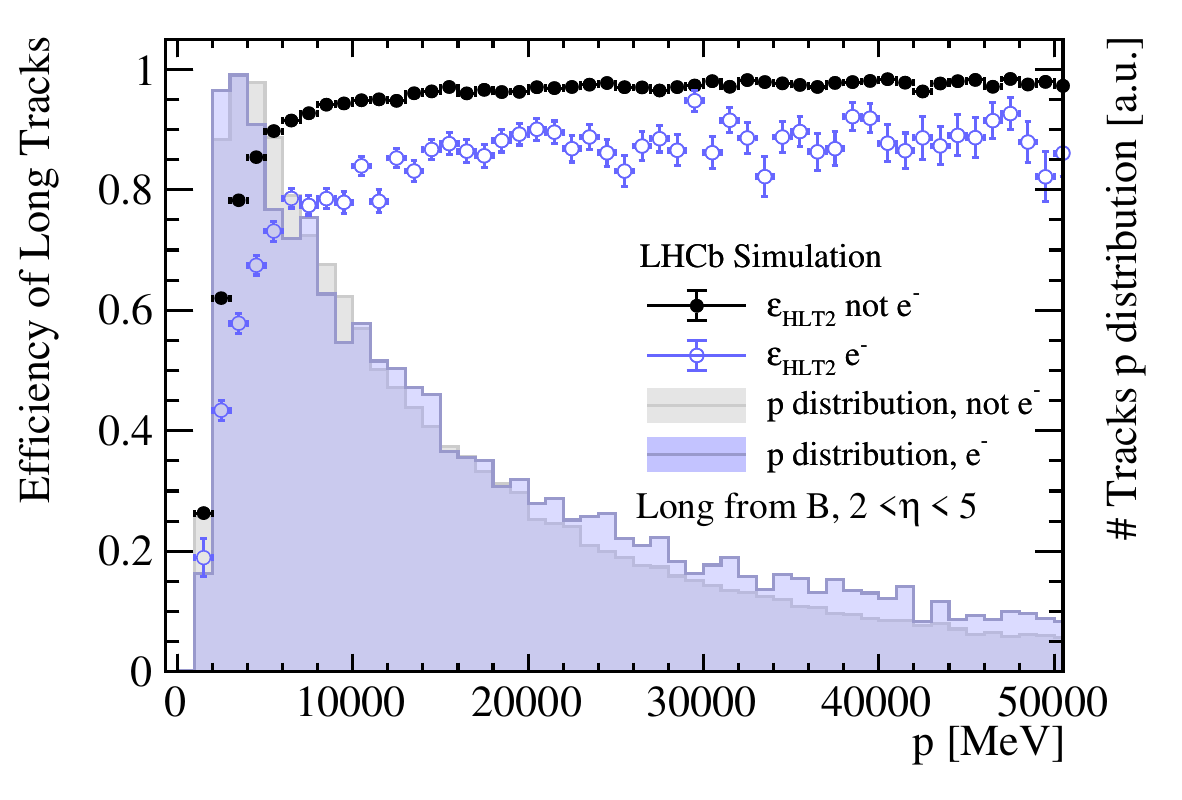}}
    \qquad
    \subfloat[]{\includegraphics[width=0.475\linewidth]{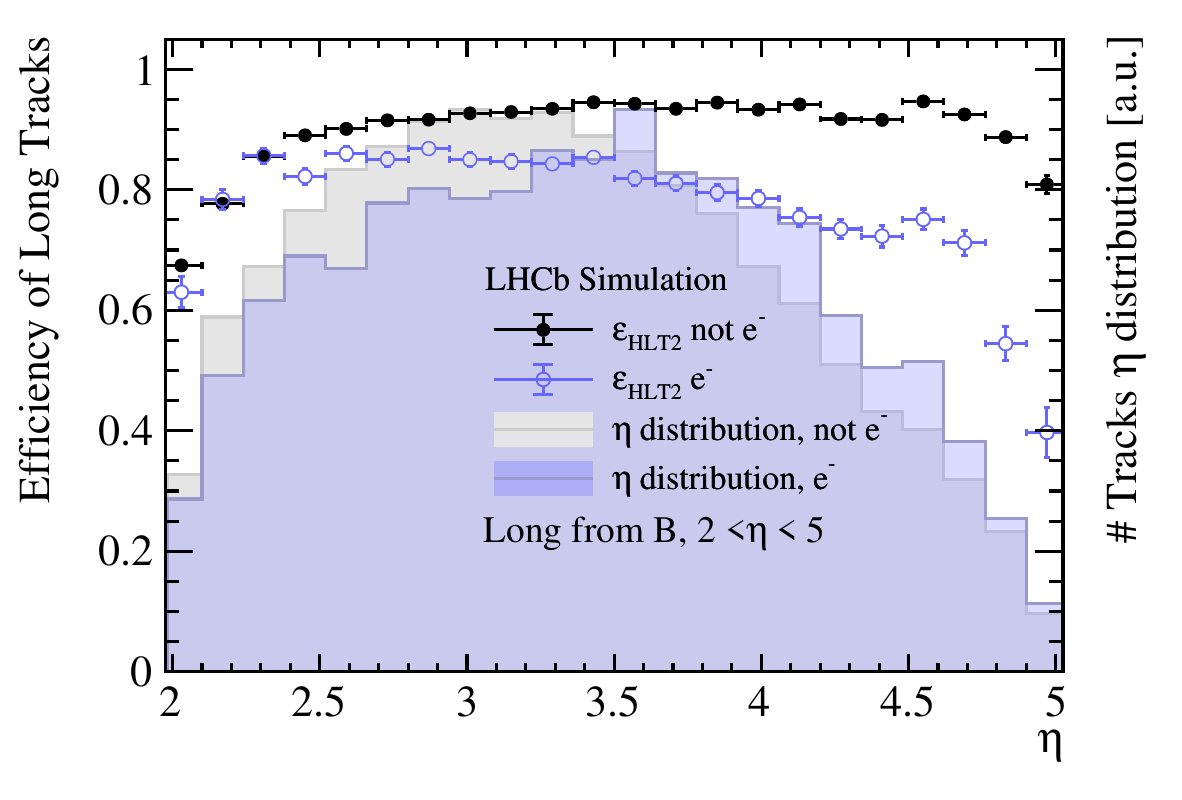}}
    \caption{Long track reconstruction efficiency of tracks reconstructed by the Forward Tracking algorithm versus momentum (a) and pseudorapidity (b) for reconstructible electrons and non-electron particles from \B decays (empty blue and filled black circles, respectively)  within $2<\eta<5$ \cite{LHCb-Figure-2022-005}. The underlying histograms show the corresponding distributions of the reconstructible particles. Being reconstructible requires a minimum number of hits in different tracking detectors as described in Ref.~\cite{Li:2021oga}.}
    \label{fig:reco_performance_eff}
\end{figure}

\begin{figure}[!htb]
    \centering
    \subfloat[]{\includegraphics[width=0.475\linewidth]{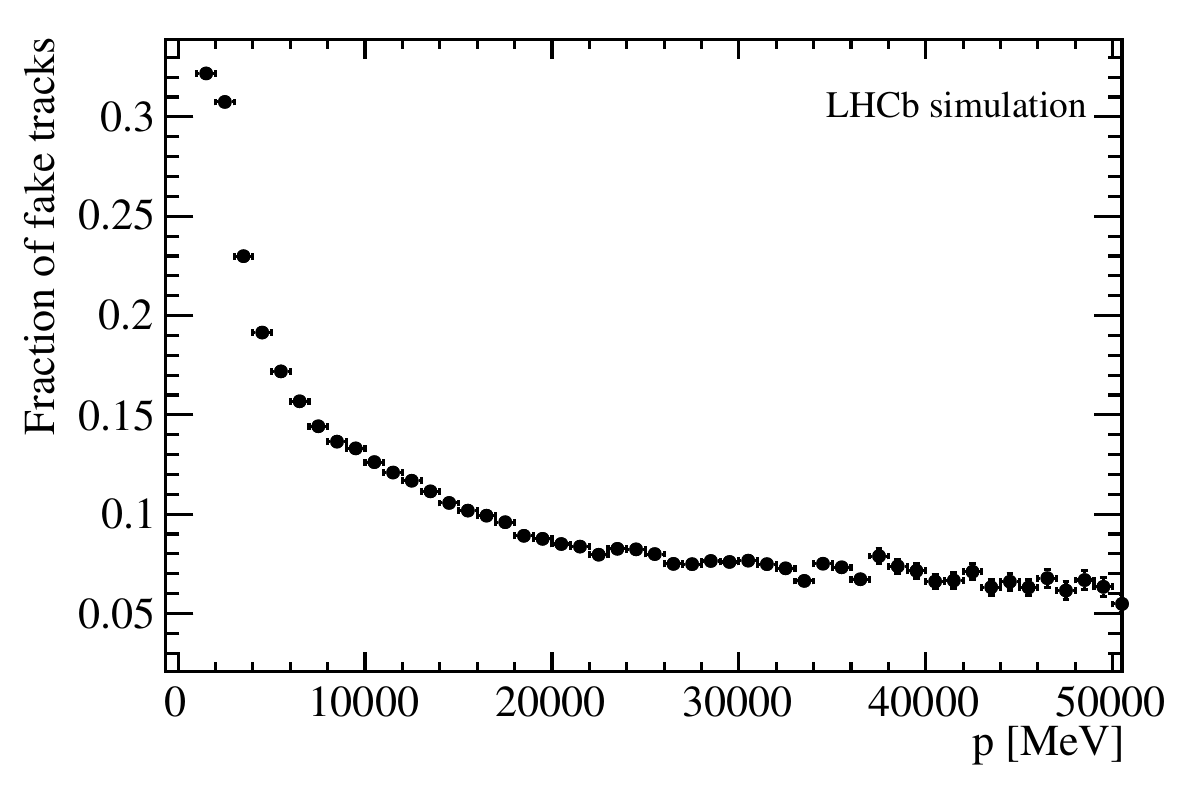}}
    \qquad
    \subfloat[]{\includegraphics[width=0.475\linewidth]{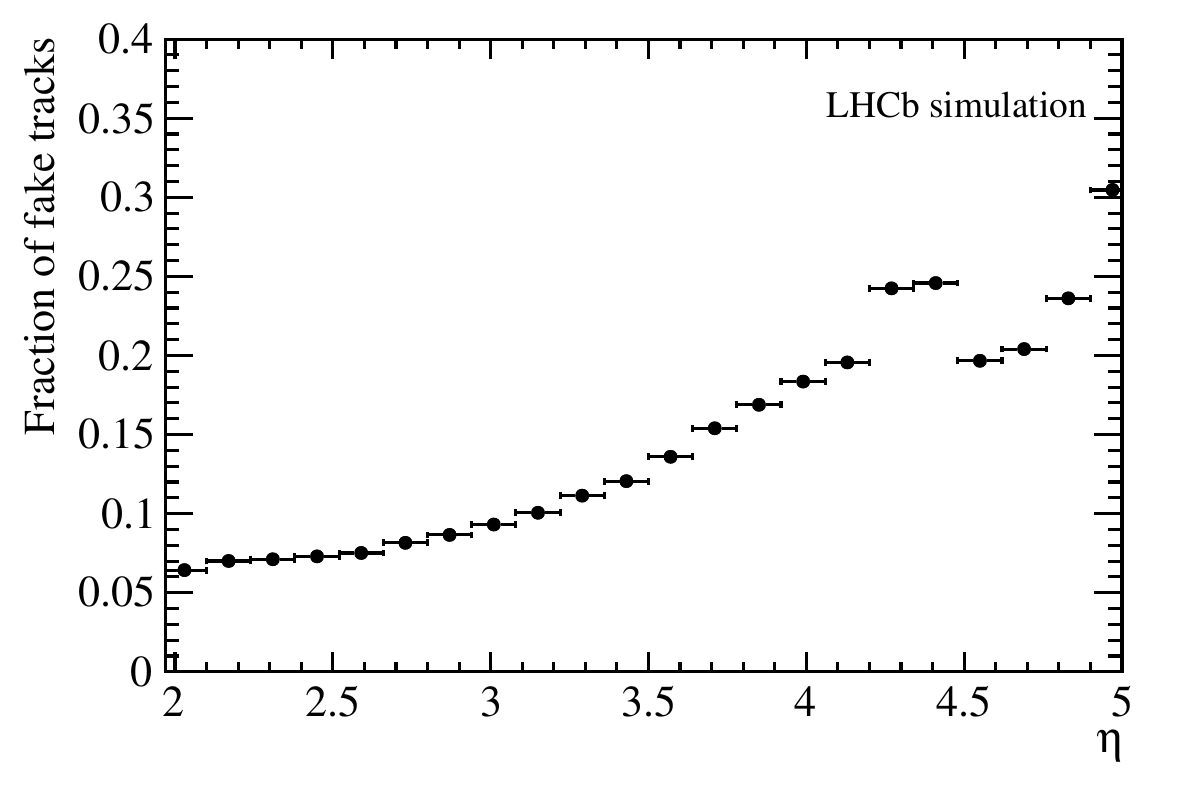}}
    \caption{Fake track fraction of Long tracks reconstructed by the Forward Tracking algorithm as a function of fake track momentum (a) and pseudorapidity (b) \cite{LHCb-Figure-2022-005}.}
    \label{fig:reco_performance_ghost}
\end{figure}

The Long track reconstruction efficiency achieved by the Forward Tracking is shown in Fig.~\ref{fig:reco_performance_eff}. A high efficiency is particularly important for studies of charm and beauty decays with many final state particles in which track reconstruction inefficiencies have a large impact on the total decay reconstruction efficiency. The integrated efficiency for reconstructing hadrons and muons originating from a \B meson is around $90$\%, while tracks with a momentum higher than $10\gev$ reach more than $95$\% efficiency. Electrons are in general harder to reconstruct with high efficiency as they lose energy predominantly via bremsstrahlung when interacting with the detector material which unpredictably alters their trajectory. This effect is particularly visible in Fig.~\ref{fig:reco_performance_eff}b in the high pseudorapidity region in which particles traverse more material.  
It is also this region that exhibits the highest fake track fraction, as shown in Fig.~\ref{fig:reco_performance_ghost}b, partly because of increased multiple scattering and hadronic interactions due to the material, but even more because of the higher SciFi hit density in the very forward direction leading to more possible random hit combinations. Fig.~\ref{fig:reco_performance_ghost}a shows the fake track fraction in dependence of the fake track momentum. The integrated fake track fraction coming from the Forward Tracking amounts to $15$\%. It is afterwards reduced by applying a Kalman Filter and evaluation of a fake track classifier. 

\section{Conclusion}

The Forward Tracking is one of the algorithms finding Long tracks in \lhcb's Run 3 \hlttwo reconstruction sequence. To avoid the redundancy in Long track finding with the Matching algorithm and hence to improve the throughput of \hlttwo, the Forward Tracking is only run on residual \velo tracks and SciFi hits. Including the Kalman filter and particle identification algorithms, the \hlttwo reconstruction sequence reaches its goal of an event throughput of more than $500$ events per second and computing node. The Forward Tracking uses a method similar to the Hough transform to recognise the line patterns left by particles traversing the SciFi detector. With this, it reaches more than $90$\% reconstruction efficiency for tracks originating from a \B meson decay and a fake track fraction of 15\%. Hence, the \hlttwo reconstruction sequence and the Forward Tracking are well prepared for data taking during Run 3 of the \lhc.
\Acknowledgements
The author would like to acknowledge the support by the \lhcb collaboration and in particular thank the \lhcb Computing, RTA and Simulation teams for their work that makes developing and benchmarking our algorithms possible.



\end{document}